\documentclass[12pt]{extarticle}
\usepackage{setspace}

\usepackage[T1]{fontenc}

\usepackage[latin1]{inputenc}
\usepackage{epsfig}
\usepackage[english,french]{babel}
\usepackage{color}
\usepackage{graphicx}

\usepackage{dcolumn}

\usepackage{moreverb}

\usepackage{amsmath,amssymb,amsfonts}

\begin{document}
\numberwithin{equation}{section}
\newcommand{\boxedeqn}[1]{%
  \[\fbox{%
      \addtolength{\linewidth}{-2\fboxsep}%
      \addtolength{\linewidth}{-2\fboxrule}%
      \begin{minipage}{\linewidth}%
      \begin{equation}#1\end{equation}%
      \end{minipage}%
    }\]%
}


\newsavebox{\fmbox}
\newenvironment{fmpage}[1]
     {\begin{lrbox}{\fmbox}\begin{minipage}{#1}}
     {\end{minipage}\end{lrbox}\fbox{\usebox{\fmbox}}}

\raggedbottom
\onecolumn

\parindent 8pt
\parskip 10pt
\baselineskip 16pt
\noindent\title*{{\LARGE{\textbf{Generalized MICZ-Kepler system, duality, polynomial and deformed oscillator algebras}}}}
\newline
\newline
Ian Marquette
\newline
Department of Mathematics, University of York, Heslington, York, UK. YO10 5DD
\newline
im553@york.ac.uk
\newline
\newline
We present the quadratic algebra of the generalized MICZ-Kepler system in three-dimensional Euclidean space $E_{3}$ and its dual the four dimensional singular oscillator in four-dimensional Euclidean space $E_{4}$. We present their realization in terms of a deformed oscillator algebra using the Daskaloyannis construction. The structure constants are in these cases function not only of the Hamiltonian but also of other integrals commuting with all generators of the quadratic algebra. We also present a new algebraic derivation of the energy spectrum of the MICZ-Kepler system on the three sphere $S^{3}$ using a quadratic algebra. These results point out also that results and explicit formula for structure functions obtained for quadratic, cubic and higher order polynomial algebras in context of two-dimensional superintegrable systems may be applied to superintegrable systems in higher dimensions with and without monopoles.
\section{Introduction}
The most well-known superintegrable systems are the Kepler-Coulomb system and the harmonic oscillator [1,2,3]. A systematic study of superintegrable systems with second-order integrals of motion (`quadratically superintegrable systems') in two-dimensional (2D) Euclidean space was begun some years ago [4,5].  For a review of 2D superintegrable systems we refer the reader to [6].

In $N$ dimensions, the symmetry algebra generated by the integrals of motion of the harmonic oscillator is the Lie algebra $su(N)$, while that of the bounded states of the hydrogen atom is the Lie algebra $so(N+1)$. A nonlinear symmetry algebra was discussed for the first time in context of the Kepler-Coulomb system and harmonic oscillator on the three sphere $S^{3}$ [7]. A quadratic algebra for these two systems and the Hartmann potential were obtained for fixed energy and angular momentum [8]. The integrals of the Hartmann system form a quadratic Hahn algebra $QH(3)$, while a quadratic Racah algebra $QR(3)$ describes the symmetry algebra of the harmonic oscillator and Kepler-Coulomb system on a 3D space of constant curvature. 

A general quadratic algebra in classical and quantum mechanics, generated by second order integrals of motion, was studied by Daskaloyannis [9], who discussed its realizations in terms of deformed oscillator algebras and found a method to obtain the energy spectrum algebraically. We generalized this method for systems with a second and a third order integral of motion, generating a cubic algebra [10]. In further work all quantum superintegrable systems separable in Cartesian coordinates with second and third order integrals were studied from the point of view of polynomial algebras [11], and a class of quintic and seventh order algebras, their realizations in terms of deformed algebras, and their finite dimensional unitary representations, were presented. Families of systems with algebras of arbitrary order were also considered. We also discussed the relations between superintegrable systems and their polynomial symmetry algebras, ladder operators and their polynomial Heisenberg algebras and supersymmetry [11].

A systematic search for superintegrable systems with magnetic fields was started in recent years [12-15]. However, the problem of degeneracy in the presence of a magnetic monopole was studied in earlier articles independently by Zwanziger [16] and by McIntosh and Cisneros  [17]. This system and his dynamical symmetry algebra was also studied in Ref.18. This `MICZ-Kepler' system, of a magnetic monopole in flat 3D space with an arbitrary electric charge plus a fixed inverse-{\em square} term in the potential, is superintegrable, and has an $so(4)$ symmetry algebra generated by the Poincar\'e vector together with a Runge-Lenz vector. The MICZ-Kepler problem also exists in higher dimensions, just as the Kepler problem does [19], where they remain superintegrable. The Yang-Coulomb system (YCS) is another superintegrable system, consisting  of the Yang monopole and a particle coupled to the monopole by isospin $su(2)$ and the Kepler-Coulomb interaction [20,21]. This system is related to the 8D harmonic oscillator and has a $so(6)$ symmetry algebra [22].

The MICZ-Kepler problem was also considered in $S^{3}$ [23]. The integrals of motion of this superintegrable system form a cubic algebra similar to that of the Kepler-Coulomb problem on $S^{3}$. Further, a generalized MICZ-Kepler system was introduced by Mardoyan [24]. This system can be seen as the intrinsic Smorodinsky-Winternitz system with monopole in 3D Euclidean space. It can be transformed into a 4D singular harmonic oscillator by a duality transformation [25]. The corresponding Schrödinger equation was solved for these two cases, but no polynomial algebras were found.

In light of these results one can ask the question if a polynomial algebra can be generated by the integrals of motion of the generalized MICZ-Kepler system and the 4D singular harmonic oscillator. The purpose of this paper is to obtain such a polynomial algebra and its realization in terms of a deformed oscillator algebra of these two systems, and to present a new derivation of their energy spectrum using an algebraic method. 

The paper is structured as follows. In Section 2, we recall results on quadratic algebras and their realization in term of deformed oscillator algebras. In Section 3 we consider the case of the generalized MICZ-Kepler system. We present the quadratic algebra, the Casimir operators, the realization in terms of a deformed oscillator algebra and its spectrum. In Section 4, we consider the case of the 4D singular oscillator. In Section 5 we present a new derivation of the energy spectrum of the MICZ-Kepler system on $S^{3}$ by constructing a quadratic algebra and using its realization in terms of deformed oscillator algebras.

\section{Quadratic algebra and realizations in term of deformed oscillator algebras}
The most general quadratic algebra [7] generated by the integrals of motion of a quadratically superintegrable system is
\[ [A,B]=C, \]
\begin{equation}
[A,C]=\beta A^{2} +\gamma\{A,B\}+\delta A+\epsilon B+\zeta,
\end{equation}
\[ [B,C]=aA^{2}-\gamma B^{2}-\beta \{A,B\}+d A-\delta B + z.\]
In the [9-11], the structure constants of the polynomial algebra were assumed to be polynomial functions of the Hamiltonian only. Such algebras were studied in the context of 2D superintegrable systems. However, because we are interested in application in context of superintegrable systems in dimension higher than two, in this paper we allow the structure constants $\beta$, $\delta$, $\gamma$, $\epsilon$, $\zeta$, $a$, $d$ and $z$ to be polynomial functions not only of the Hamiltonian but also of any other integrals of motion ($F_{i}$) that commute with the Hamiltonian, with each other and also with the generators of the quadratic algebra $A$, $B$ and $C$ (that is, $[F_{i},H]=[F_{i},F_{j}]=[F_{i},A]=[F_{i},B]=[F_{i},C]=0$). When we study this algebra's realization in terms of deformed oscillator algebras and its representations, we will fixed the energy ($H\psi=E\psi$) but also these other integrals ($F_{i}\psi=f_{i}\psi$) of motion that form, with the Hamiltonian, an Abelian subalgebra. The Casimir operator ($[K,A]=[K,B]=[K,C]=0$) of this quadratic algebra is thus given in terms of the generators by
\begin{equation}
K=C^{2}-\alpha\{A^{2},B\}-\gamma\{A,B^{2}\}+(\alpha \gamma -\delta)\{A,B\}+(\gamma^{2}-\epsilon)B^{2}
\end{equation}
\[+(\gamma \delta -2 \zeta)B+\frac{2a}{3}A^{3}+(d+\frac{a\gamma}{3}+\alpha^{2})A^{2}+(\frac{a\epsilon}{3}+\alpha\delta+2z)A.\]
The  Casimir operator $K$ will be also rewritten as a polynomial of $H$ and $F_{i}$. 

The realizations of the quadratic algebra in terms of deformed oscillator algebras were obtained in [9]:
\begin{equation}
[N,b^{\dagger}]=b^{\dagger},\quad [N,b]=-b,bb^{\dagger}=\Phi(N+1),b^{\dagger}b=\Phi(N),
\end{equation}
and the structure function for the case $\epsilon=\delta=\beta=0$ is given by 
\begin{equation}
\hspace*{-0.4in}\Phi(N)= -3072 \gamma^{6}K(-1+2(N+u))^{2}-48\gamma^{6}(-d\gamma^{2})(-3+2(N+u))(-1+2(N+u))^{4}(1+2(N+u))
\end{equation}
\[\hspace*{-0.2in}+\gamma^{8}(4a\gamma)( -3+2(N+u))^{2}(-1+2(N+u))^{4}(1+2(N+u))^{2}+ 768(4\gamma^{2}\zeta)^{2}+ 32\gamma^{4}(-1+2(N+u))^{2}\]
\[+(-1-12(N+u)+12(N+u)^{2})(8 \gamma^{3}z)-256\gamma^{2}(-1+2(N+u))^{2}(-4\gamma^{5}z).\]
(Such realizations for cubic and higher order polynomial algebras have  also been obtained [10,11].) 

To obtain unitary representations we should impose the following three constraints on the structure function:
\begin{equation}
\Phi(p+1,u,k)=0, \quad \Phi(0,u,k)=0,\quad \phi(x)>0, \quad \forall \; x>0 \quad .
\end{equation}
The energy $E$ and the arbitrary constant $u$ are solutions of the equation obtained from these constraints.
\section{Generalized MICZ-Kepler systems}
The generalized MICZ-Kepler system was introduced and studied by Mardoyan [24], where it was shown that the Schrödinger equation for this system allows separation of variables in spherical and parabolic coordinates, and the spectral problem was solved. We use the system of units for which ( $\hbar=m=e=c=1$ )

The Hamiltonian of the generalized MICZ-Kepler system is given by
\begin{equation}
H=\frac{1}{2}(-i\nabla-s \vec{A})^{2}+\frac{s^{2}}{2r^{2}}-\frac{1}{r}+\frac{c_{1}}{r(r+z)}+\frac{c_{2}}{r(r-z)},
\end{equation}
with
\begin{equation}
A=\frac{1}{r(r-z)}(y,-x,0).
\end{equation}
Let us recall the Poincar\'e and  Laplace-Runge-Lenz vectors for the MICZ-Kepler system on the three-dimensional Euclidean space $E_{3}$:
\begin{equation}
\vec{J}=\vec{r}\times (-i\nabla-s\vec{A})-s\frac{\vec{r}}{r},
\end{equation}
\begin{equation}
\vec{I'}=\frac{1}{2}(\vec{J}\times (-i\nabla-s\vec{A})-(-i\nabla-s\vec{A})\times \vec{J}+\frac{\vec{r}}{r}.
\end{equation}
The generalized MICZ-Kepler system has the following integrals of motion ($[H,A]=[H,B]=[H,J_{z}]=0$):
\begin{equation}
A=\vec{J}^{2}+\frac{2rc_{1}}{(r+z)}+\frac{2rc_{2}}{(r-z)},\quad B=I'_{z}+\frac{c_{1}(r-z)}{r(r+z)}-\frac{c_{2}(r+z)}{r(r-z)},\quad J_{z}.
\end{equation}
From these, we can form the  quadratic algebra (with $[A,J_{z}]=[B,J_{z}]=0$)
\[  [A,B]=C, \]
\begin{equation}
[A,C]=2\{A,B\}+4sJ_{z}+4(c_{2}-c_{1}),
\end{equation}
\[ [B,C]=-2B^{2}+8HA-4J_{z}^{2}H+4(1-s^{2}-2c_{1}-2c_{2})H+2, \]
which is the Daskaloyannis quadratic algebra at the special values of its structure constants
\begin{equation}
\beta=\delta=\epsilon=a=0,\quad \gamma=2,\quad d=8H,\quad \zeta=4sJ_{z}+4(d-c),
\end{equation}
\[ z=(-4J_{z}^{2}-4s^{2}-4(2c_{1}+2c_{2}-1))H+2 . \]

The Casimir operator can be rewritten in terms of the Hamiltonian and $J_{z}$ (which commute with all generators of the quadratic algebra) as
\begin{equation}
K=-8s^{2}J_{z}^{2}H+16(c-d)sJ_{z}H-8(c_{1}-c_{2})^{2}H+4J_{z}^{2}+4(2c_{1}+2c_{2}+s^{2}).
\end{equation}
Using (2.4), (3.7) and (3.8) (with the energy and the $z$ component of the Poincar\'e vector taking the values $H\psi=E\psi$ and $J_{z}\psi=m\psi$) we then obtain the  structure function (with $m_{1}=\sqrt{(m-s)^2+4c_{1}}$, $m_{2}=\sqrt{(m+s)^2+4c_{2}}$)
\begin{equation}
\hspace*{-0.3in}\Phi(x)=-3\cdot 2^{21}E(x+u-(\frac{1}{2}+\frac{1}{\sqrt{-2E}})) (x+u-(\frac{1}{2}-\frac{1}{\sqrt{-2E}}))
\end{equation}
\[(x+u-(\frac{1}{2}-(\frac{m_{1}+m_{2}}{2})))(x+u-(\frac{1}{2}-(\frac{m_{1}-m_{2}}{2})))\]
\[(x+u-(\frac{1}{2}-(\frac{-m_{1}+m_{2}}{2})))(x+u-(\frac{1}{2}-(\frac{-m_{1}-m_{2}}{2})))\]
From the constraint (2.5) we then obtain the following finite dimensional unitary representation and energy spectrum
\begin{equation}
E=\frac{-1}{2(p+1+m_{1}+m_{2})^{2}},\quad u=\frac{1}{2}+\frac{1}{\sqrt{-2E}},
\end{equation}
\begin{equation}
\Phi(x)=3\cdot 2^{21}\frac{x(p+1-x)}{(p+1+m_{1}+m_{2})^{2}}(p+1+m_{1}-x)(p+1+m_{2}-x)
\end{equation}
\[(p+1+m_{1}+m_{2}-x)(2p+2+m_{1}+m_{2}-x).\]
In the Ref.23 the energy spectrum in terms of the principal quantum number was presented
\begin{equation}
E=E_{n}^{(s)}=\frac{-1}{2(n+\frac{\delta_{1}+\delta_{2}}{2})^{2}}
\end{equation}
The Eq.(3.9) of the Ref.24 relate the parabolic quantum numbers $n_{1}$ and $n_{2}$ with the principal quantum number $n$ 
\begin{equation}
n=n_{1}+n_{2}+\frac{|m-s|+|m+s|}{2}+1,\quad m_{1}=|m-s|+\delta_{1},\quad m_{2}=|m+s|+\delta_{2}
\end{equation}
By taking $p=n_{1}+n_{2}$ the formula given by Eq.(3.10) and Eq.(3.12) coincide.

To summarize: We have obtained the quadratic algebra for the generalized MICZ-Kepler system, constructed its realization in term of deformed oscillator algebra, and obtained the energy spectrum. This is the first study of this system from the point of view of its polynomial algebra.
\section{Four dimensional singular oscillator}
A relation between systems with and without monopoles has been discussed in [22], and such a relation was obtained for the generalized MICZ-Kepler system system and a 4D singular harmonic oscillator [25,26], which is also superintegrable. Let us recall results concerning these systems and the  duality transformation. We consider the following change of variables [25]:
\begin{equation}
\psi^{(s)}(\vec{r})\mapsto \psi(\vec{r},\gamma)=\psi^{(s)}(\vec{r})\frac{e^{is(\gamma-\phi)}}{\sqrt{4\pi}},\quad s \mapsto -i\frac{\partial}{\partial \gamma}, \quad \gamma \in [0,4\pi).
\end{equation}
The generalized MICZ-Kepler system is then  transformed into a 4D double singular oscillator [26]. If we change $\beta=\theta$ and $\alpha=\phi$ and pass from the coordinates $r,\alpha,\beta,\gamma$ to the coordinates
\begin{equation}
u_{0}+iu_{1}=u\cos(\frac{\beta}{2})e^{i\frac{\alpha+\gamma}{2}},\quad u_{2}+iu_{3}=u\sin(\frac{\beta}{2})e^{i\frac{\alpha-\gamma}{2}},
\end{equation}
we obtain the Hamiltonian
\begin{equation}
H=-\frac{1}{2}\left(\frac{\partial^{2}}{\partial u_{0}^{2}}+\frac{\partial^{2}}{\partial u_{1}^{2}}+\frac{\partial^{2}}{\partial u_{2}^{2}}+\frac{\partial^{2}}{\partial u_{3}^{2}}\right)+\frac{ \omega^{2} u^{2}}{2}+\frac{c_{1}}{u_{0}^{2}+u_{1}^{2}}+\frac{c_{2}}{u_{2}^{2}+u_{3}^{2}}.
\end{equation}
This system was studied in  [26], and is separable in Eulerian, double-polar and spheroidal coordinates in $E_{4}$. It has the following integrals of motion ( $[H,J_{1}]=[H,J_{2}]=[H,A,]=[H,B]=0$):
\begin{equation}
J_{1}=-i(u_{0}\frac{\partial}{\partial u_{1}}-u_{1}\frac{\partial}{\partial u_{0}}),\quad J_{2}=-i(u_{2}\frac{\partial}{\partial u_{3}}-u_{3}\frac{\partial}{\partial u_{2}}),
\end{equation}
\begin{equation}
A=-\frac{1}{4} \left(u^{2}\frac{\partial^{2}}{\partial u_{i}^{2}}-u_{i}u_{j}\frac{\partial^{2}}{\partial u_{i}\partial u_{j}}-3u_{i}\frac{\partial}{\partial u_{i}}\right)+\frac{ u^{2}}{2}\left(\frac{c_{1}}{u_{0}^{2}+u_{1}}+\frac{c_{2}}{u_{2}+u_{3}}\right),
\end{equation}
\begin{equation}
B=-\frac{1}{2}\left(\frac{\partial^{2}}{\partial u_{0}^{2}}+\frac{\partial^{2}}{\partial u_{1}^{2}}-\frac{\partial^{2}}{\partial u_{2}^{2}}-\frac{\partial^{2}}{\partial u_{3}^{2}}\right)+\frac{ \omega^{2} }{2}(u_{0}^{2}+u_{1}^{2}-u_{2}^{2}-u_{3}^{2})
\end{equation}
\[+\frac{c_{1}}{u_{0}^{2}+u_{1}^{2}}-\frac{c_{2}}{u_{2}^{2}+u_{3}^{2}}.\]

We then have the  polynomial algebra ( with $J_{3}=-\frac{1}{2}J_{1}-\frac{1}{2}J_{2}$ and $J_{3}'=-\frac{1}{2}J_{1}+\frac{1}{2}J_{2}$ and $[A,J_{3}]=[A,J_{3}']=[B,J_{3}]=[B,J_{3}']=0$)
\[ [A,B]=C,\]
\begin{equation}
[A,C]=2\{A,B\}+4J_{3}J'_{3}H-2(c_{1}-c_{2})H,
\end{equation}
\[ [B,C]=-2B^{2}-16\hbar^{2}\omega^{2}A+2H^{2} -8\omega^{2}J_{3}^{2}-8\omega^{2}J_{3}^{'2}+8(c_{1}+c_{2}-1)\omega^{2},\]
for which the structure constants may be obtained by comparing (2.1) and (4.7):
\begin{equation}
\beta=a=\delta=\epsilon=0,\quad \gamma=2,\quad d=-16a^{2}
\end{equation}
\[ \zeta=4J_{3}J'_{3}H-2(c_{1}-c_{2})H,\quad z=2H^{2}-8\omega^{2}(J_{3}^{2}+J_{3}^{'2})+8(c_{1}+c_{2}-1)\omega^{2}\]
We can obtain the Casimir operator and rewrite it in terms of $H,J_{3},J_{3}'$ (which commute with all other integrals of motion):
\begin{equation}
K=-4J_{3}^{2}H^{2}-4J_{3}'^{2}H^{2}+4(c_{1}+c_{2})H^{2}+4\omega^{2}J_{3}^{2}J_{3}^{'2}
\end{equation}
\[-16(c_{1}-c_{2})\omega^{2}J_{3}J_{3}'+4(c_{1}-c_{2})^{2}\omega^{2}.\]
The structure constants (4.8) and the Casimir operator (4.9) allow us to obtain the structure function $\Phi(x)$ of the deformed oscillator algebra. The energy and the two angular momenta are fixed, $H\psi=E\psi$, $J_{3}\psi=m\psi$ and $J_{3}'\psi=s\psi$. The structure function is then given by (with $m_{1}=(m+s)^{2}+2c_{1}$ and $m_{2}=(m-s)^{2}+2c_{2}$)
\begin{equation}
\Phi(x)=3\cdot 2^{22}\omega^{2}\left(x+u-\left(\frac{1}{2}-\frac{E}{2\omega}\right)\right)\left(x+u-\left(\frac{1}{2}+\frac{E}{2\omega}\right)\right)
\end{equation}
\[ \left(x+u-\left(\frac{1}{2}-\frac{1}{2}(m_{1}+m_{2})  \right)\right) \left(x+u-\left(\frac{1}{2}-\frac{1}{2}(m_{1}-m_{2})  \right)\right)\]
\[ \left(x+u-\left(\frac{1}{2}-\frac{1}{2}(-m_{1}+m_{2})  \right)\right) \left(x+u-\left(\frac{1}{2}-\frac{1}{2}(-m_{1}-m_{2})  \right)\right),\]
We found the following finite dimensional unitary representation:
\begin{equation}
E=2\omega(p+1+\frac{m_{1}+m_{2}}{2}),\quad u=\frac{1}{2}-\frac{E}{2\omega},
\end{equation}
\begin{equation}
\phi(x)=3\cdot 2^{19}\omega^{2}x(p+1-x)(p+1+m_{1}-x)(p+1+m_{2}-x)
\end{equation}
\[(p+1+m_{1}+m_{2}-x)(2p+2+m_{1}+m_{2}-x)  .\]
In Ref.26 the systems was studied using double polar coordinates and the double polar quantum numbers $n_{1}$ and $n_{2}$. There are relations similar as Eq.(3.13). These results coincide by taking $p=n_{1}+n_{2}$. 

To summarize: We have presented an algebraic derivation of the energy spectrum of this superintegrable system, and its finite-dimensional unitary representations. This system is interesting not only for the duality (4.1) but also because 4D systems have not been studied systematically, and the application to them of polynomial algebra, with its power to determine spectra, is an unexplored subject. We leave to a future article the study of the corresponding classical superintegrable system and the polynomial Poisson algebra. 
\section{MICZ-Kepler system on $S^{3}$, deformed $so(4)$ algebras and quadratic algebra}
The quantum mechanical problem of motion of a arbitrary dyon field modified by a centrifugal term proportional to the square of the magnetic charge, (the MICZ-Kepler problem) has also been considered in the 3D space of constant positive curvature $S^{3}$ [23]. The integrals of motion of this superintegrable system form a cubic algebra similar to that of the Kepler-Coulomb problem on $S^{3}$. The explicit form of this cubic algebra may be used to obtain the energy spectrum of the problem [23]. We will point out that we can form a quadratic algebra with three generators and use results obtained in context of 2D systems to obtain the energy spectrum. 

We begin by recalling some known results concerning the MICZ-Kepler system on $S^{3}$.
In coordinates $\textbf{x}=\{x_{1},x_{2},x_{3}\}$ and $x_{\mu}x^{\mu}=\textbf{x}^{2}+x_{4}^{2}=R^{2}$, the metric of the sphere is $ds^{2}=(dx)^{2}+\frac{(xdx)^{2}}{R^{2}-x^{2}}$ and the Hamiltonian
\begin{equation}
H_{\mu}=-\frac{1}{2}(\nabla^{a}+iA^{a})(\nabla_{a}+iA_{a})+\frac{\mu^{2}x_{4}^{2}}{2R^{2}|x|^{2}}-\frac{\alpha x_{4}}{R|x|},
\end{equation}
\begin{equation}
A_{1}=\mu\frac{-x_{2}}{|x|(|x|+x_{3})},\quad A_{2}=\mu\frac{x_{1}}{|x|(|x|+x_{3})},\quad A_{3}=0.
\end{equation}
From [22] we then have
\begin{equation}
\pi_{a}=-i\partial_{a}+A_{a},\quad p_{4}=-i\partial_{4},\quad N_{a}=x_{4}\pi_{a}-x_{a}p_{4},
\end{equation}
with
\begin{equation}
[\pi_{a},x_{b}]=-i\delta_{ab},[\pi_{a},p_{4}]=0,[\pi_{a},\pi_{b}]=
i\mu\epsilon_{abc}\frac{x_{c}}{|x|^{3}},[p_{4},x_{4}]=-i,
\end{equation}
\begin{equation}
[N_{a},N_{b}]=i\epsilon_{abc}J_{c}+R^{2}F_{ab},\quad [J_{a},N_{b}]=i\epsilon_{abc}N_{c},
\end{equation}
\[ [J_{a},J_{b}]=i\epsilon_{abc}J_{c},F_{ab}=[\pi_{a},\pi_{b}]. \]
The integrals of motion are the Poincar\'e and Laplace-Runge-Lenz vector
\newline
\begin{equation}
J_{a}=\epsilon_{abc}x_{b}\pi_{c}-\frac{\mu x_{a}}{|x|},\quad A_{a}=\frac{1}{2R}\epsilon_{abc}(J_{b}N_{c}-N_{b}J_{c})+\frac{\alpha x_{a}}{|x|}.
\end{equation}
The MICZ-Kepler Hamiltonian on the sphere can be rewritten
\begin{equation}
H=\frac{J^{2}+N^{2}}{2R^{2}}+\frac{\mu^{2}x_{4}^{2}}{2R^{2}|x|^{2}}-\frac{\alpha x_{4}}{R|x|},
\end{equation}
and the cubic algebra [21] is then given by
\begin{equation}
[A_{a},A_{b}]=-2i(H-\frac{J^{2}}{R^{2}}+\frac{\mu^{2}}{2R^{2}})\epsilon_{abc}J_{c},
\end{equation}
\[ [J_{a},A_{b}]=i\epsilon_{abc}A_{c},\quad [J_{a},J_{b}]=i\epsilon_{abc}J_{c} \]
with the relations
\begin{equation}
\vec{A}^{2}=2H(J^{2}-\mu^{2}+1)-\frac{1}{R^{2}}J^{2}(J^{2}-\mu^{2}+2)+\alpha^{2},
\end{equation}
\begin{equation}
\vec{A}\cdot \vec{J}=\vec{J}\cdot \vec{A}=-\alpha\mu.
\end{equation}
\subsection{Quadratic algebra and realizations in term of deformed oscillator algebras}
The representations of this algebra were discussed in terms of the cubic algebra [22] given by (5.8). Deformations of the Lie algebra $so(4)$, $so(3,1)$ and $e(3)$ that leave their $so(3)$ subalgebra undeformed and preserve their coset structure were studied by Quesne [27] and the cubic algebra (5.8) is a particular case. We will point out how results obtained for polynomial algebra in the context of 2D superintegrable systems can be used to obtain the finite-dimensional unitary representation and the corresponding degenerate energy spectrum. The duality transformation for the Kepler-coulomb system in 2,3 and 5 dimensional sphere was investigated in Ref. 28. Let us consider the three generators
\begin{equation}
A=J^{2},\quad B=A_{3},\quad C=[A,B],
\end{equation}
and form the quadratic algebra
\begin{equation}
[A,B]=C,\quad [A,C]=2\{A,B\}+4\alpha \mu J_{z},
\end{equation}
\[ [B,C]=-2B^{2}-\frac{6}{R^{2}}A^{2}+(8H+\frac{4}{R^{2}}(J_{z}^{2}+\mu^{2}-1))A \]
\[+(2\alpha^{2}+4H(1-J_{z}^{2}-\mu^{2})-2\frac{J_{z}^{2}\mu^{2}}{R^{2}}) .\]
By comparing (2.4) and (5.12) we obtain the structure constants
\begin{equation}
\beta=\delta=\epsilon=0,\quad \gamma=2,\quad \zeta=4\alpha\mu J_{z},\quad a=-\frac{6}{R^{2}},
\end{equation}
\[d=8H+\frac{4}{R^{2}}(J_{z}^{2}+\mu^{2}-1),\quad z=2\alpha^{2}+4H(1-J_{z}^{2}-\mu^{2})-2\frac{J_{z}^{2}\mu^{2}}{R^{2}}.    \]
The structure constants can be used to obtain the Casimir operator $K$ of the quadratic algebra. We rewrite this operator as a polynomial of the Hamiltonian and $J_{z}$ using (2.2) and the two Casimir operators of the quadratic algebra (5.9) and (5.10):
\begin{equation}
K=4\alpha^{2}\mu^{2}+4J_{z}^{2}\alpha^{2}-8\frac{J_{z}^{2}\mu^{2}}{2R^{2}}-8J_{z}^{2}\mu^{2}H.
\end{equation}
The energy and the $z$-component of the Poincar\'e vector $J_{z}$ are again fixed, $H=E$ and $J_{z}=m$. For the MICZ-Kepler systems on $S^{3}$ we
then use (2.4), (5.13) and (5.14) to obtain the  structure function 
\begin{equation}
\Phi(x)=-\frac{12288}{R^{2}}(4m^{2}-(1-2(x+u))^{2})(-16R^{2}\alpha^{2}-8ER^{2}(1-2(x+u))^{2}+(1-2(x+u))^{2}
\end{equation}
\[+(1-2(x+u))^{2}(-3-4(x+u)+4(x+u)^{2}))(1-4(x+u)+4(x+u)^{2}-4\mu^{2}).\]
This can be rewritten in the following form (with $E'=1+2ER^{2}$):
\[\hspace*{-0.2in}\Phi(x)=-\frac{32^{18}}{R^{2}}(x+u-(\frac{1}{2}(1-2m)))(x+u-(\frac{1}{2}(1+2m)))(x+u-(\frac{1}{2}(1-2\mu)))(x+u-(\frac{1}{2}(1+2\mu)))\]
\[(x+u-(\frac{1}{2}-\frac{\sqrt{2}}{2}\sqrt{E'-\sqrt{(E')^{2}+4R^{2}\alpha^{2}}}))(x+u-(\frac{1}{2}+\frac{\sqrt{2}}{2}\sqrt{E'-\sqrt{(E')^{2}+4R^{2}\alpha^{2}}}))\]
\[(x+u-(\frac{1}{2}-\frac{\sqrt{2}}{2}\sqrt{E'+\sqrt{(E')^{2}+4R^{2}\alpha^{2}}}))(x+u-(\frac{1}{2}+\frac{\sqrt{2}}{2}\sqrt{E'+\sqrt{(E')^{2}+4R^{2}\alpha^{2}}})).\]
We obtain, with the constraint given by (2.19),
\begin{equation}
E=-\frac{\alpha^{2}}{2N^{2}}+\frac{N^{2}-1}{2R^{2}},\quad u=\frac{1}{2}(1+2|\mu|),\quad N=p+1+|\mu|,
\end{equation}
\begin{equation}
\hspace*{-0.4in}\Phi(x)=\frac{32^{18}}{R^{2}}x(p+1-x)(x+2|\mu|)(x+m+|\mu|)(x-m+|\mu|)(p+1+x+|\mu|)((2x+2|\mu|)^{2}+\frac{R^{2}\alpha^{2}}{(p+1+|\mu|)^{2}}).
\end{equation}
 To summarize: We have obtained the finite dimensional unitary representations and the corresponding energy spectrum, and have thus obtained a new algebraic derivation of the energy spectrum of this system using algebraic results obtained in the context of 2D superintegrable systems.
\newline
\section{Conclusion}
In this paper we considered the generalized MICZ-Kepler systems and the 4D singular harmonic oscillator. These systems are respectively in the 3D and the 4D Euclidean space. We presented the quadratic algebra for these two systems (for which, to our knowledge, no polynomial algebras have been obtained before), their realizations in terms of deformed oscillator algebras, their finite dimensional unitary representations and energy spectrum. The use of polynomial algebras to solve superintegrable systems in dimensions higher than two is a relatively unexplored subject [29], and these systems, with the duality relation between them, provide an important example. 

We pointed out also that results obtained in [7-9] in the context of 2D superintegrability may be applied to other superintegrable systems in higher dimensions. The structure constants of the quadratic algebras and the structure function involve not only the Hamiltonian and but also other involutive integrals. In the case of the generalized MICZ-Kepler system there is one other such integral of motion ($J_{z}$) and in the case of the 4D singular oscillator there are two such integrals ($J_{1}$ and $J_{2}$).

We also obtained a quadratic algebra for the MICZ-Kepler systems on $S^{3}$ from the cubic algebra with six generators and two Casimir operators. We used the results of Daskaloyannis [9] to obtain realizations of this in terms of deformed oscillator algebras and presented an alternative algebraic derivation of the energy spectrum. Thus the problem of finding the finite dimensional unitary representations and the corresponding energy spectrum for this system can be achieved by applying the constraint (2.5) to the structure function and solving an algebraic system of equation.

Similarly as for the Kepler-Coulomb systems, the Hartman potential and the Kepler-Coulomb system and harmonic oscillator on the three sphere $S^{3}$ [7,8] the the MICZ-Kepler systems on $S^{3}$, the generalized MICZ-Kepler systems and the 4D sungular oscillator possess a quadratic algebra with three generators and we can reobtain the energy spectrum from this algebra. However, results obtained in context of two-dimensional systems should be applied carefully in context of higher dimensional systems. A systematic study of higher dimensional suprintegrable systems and their polynomial algebras remains to be done.

A systematic search for systems with spin was also started recently [30]. It could be interesting to see how results obtained in [7-9] could be applied in this context.

The MICZ-Kpler systems [31-32], the MICZ-Kepler systems on the three sphere [33] and the generalized Kepler system [34] were discussed from the point of view of supersymmetric quantum mechanics. The factorization of the generalized MICZ-Kepler systems were also discussed [35]. We pointed out in context of two-dimensional superintegrable systems a relation between higher order integrals of motion, higher order ladder operators and higher order supersymmetric quantum mechanics [10,11]. Such relation should be to be more studied in context of systems with and without monopole.

\textbf{Acknowledgments} We thank Niall MacKay for discussions and a careful reading of the manuscript. The research of I.M. was supported by a postdoctoral
research fellowship from FQRNT of Quebec.

\section{\textbf{References}}

1. V.Fock, Z.Phys. 98, 145-154 (1935).
\newline
2. V.Bargmann, Z.Phys. 99, 576-582 (1936).
\newline
3. J.M.Jauch and E.L.Hill, Phys.Rev. 57, 641-645 (1940).
\newline
4. J.Fris, V.Mandrosov, Ya.A.Smorodinsky, M.Uhlir and P.Winternitz,  Phys.Lett. 16, 354-356 (1965).
\newline
5. P.Winternitz, Ya.A.Smorodinsky, M.Uhlir and I.Fris, Yad.Fiz. 4,
625-635 (1966). (English translation in Sov. J.Nucl.Phys. 4,
444-450 (1967)).
\newline
6. I.Marquette and P.Winternitz, J. Phys. A: Math. Theor. 41, 304031 (2008).
\newline
7. P.W.Higgs, J.Phys. A: Math. Gen. 12 (3) 309 (1979), H.I.Leemon, J.Phys. A: Math. Gen. 12 (4) 489 (1979).
\newline
8. Ya.I.Granovskii, A.S.Zhedanov and I.M.Lutzenko, J.Phys. A: Math. Gen. 24 3887 (1991). Ya.I.Granovskii, A.S.Zhedanov and I.M.Lutzenko, Theoret. and Math. Phys. 89 474-480 (1992), Theoret. and Math. Phys. 91  604-612 (1992).
\newline
9. C.Daskaloyannis, J.Math.Phys. 42, 1100 (2001).
\newline
10. I.Marquette, J. Math. Phys. 50, 012101 (2009); J. Math. Phys. 50 095202 (2009).
\newline
11. I.Marquette, J.Math.Phys. 50 122102 (2009); I.Marquette, J.Phys A: Math. Gen. 43 135203 (2010).
\newline
12. B.Dorizzi, B.Grammaticos, A.Ramani and P.Winternitz,
J.Math.Phys. 26, 3070-3079 (1985).
\newline
13. E.McSween and P.Winternitz, J.Math.Phys. 41, 2957-2967 (2000).
\newline
14. J.B\'erub\'e and P.Winternitz, J.Math.Phys. 45, 1959-1973
(2004).
\newline
15. F.Charest, C.Hudon and P.Winternitz, J.Math.Phys. 48,
012105.1-16 (2007).
\newline
16. H.V.McIntosh and A.Cisneros, J.Math.Phys. 11 3 (1970).
\newline
17. D.Zwanziger, Phys. Rev. 176 5 (1968).
\newline
18. R.Jakiw, Ann.Phys. (NY) 129 183 (1980).
\newline
19. G.Meng, J.Phys. Math. 48, 032105 (2007).
\newline
20. C.N.Yang, J.Math. Phys. 19 320 (1978).
\newline
21. L.G.Mardoyan, a.N.Sissukian and U.M.Ter-Antonyan, Hidden Symmetry of the Yang-Coulomb System 9803010.
\newline
22. L.G.Mardoyan, A.N.Sissakian and V.M.Ter-Antonyan, Mod.Phys. Lett. A 14 1303 (1999); A.Nersessian, G.Pogosyan, Phys. Rev. A 63 020103 (2001); V.H.Le, T.-S.Nguyen and N.-H.Phan, J.Phys. A:Math. Theor. 42 175204 (2009).
\newline
23. V.V.Gritsev, Yu.A.Kurochkin and V.S.Otchik, J.Phys.A: Math. Gen 33 4903-4910 (2000).
\newline
24. L.Mardoyan, J.Math.Phys. 44 11 (2003).
\newline
25. L.G.Mardoyan and M.G.Petrosyan, Phys. Atom. Nucl. 70 572-575 (2007).
\newline
26. M.Petrosyan, Phys. Atom. Nucl. 71 1094-1101 (2008).
\newline
27. C.Quesne, J.Phys. A: Math. Gen 28 2847-2860 (1995).
\newline
28. E.G.Kalnins, W.Miller Jr. and G.S.Pogosyan, J.Math.Phys. 41, 2629-2657 (2000).
\newline
29. Y. Tanoudis, C. Daskaloyannis, XXVII Colloquium on Group Theoretical Methods in Physics, Yerevan, Armenia, Aug. (2008) arXiv:0902.0130 ; 4th Workshop on Group Analysis of Differential Equations and Integrable Systems, Protaras, Cyprus, Oct. (2008)  arXiv:0902.0259.
\newline
30. Pavel Winternitz and Ysmet Yurdusen, J.Math. Phys. 47, 103509 2006; J.Phys. A: Math. Gen. 42 385203 (2009)
\newline
31. E.D'Hoker and L.Vinet, Commun. Math.Phys. 97, 391-427 (1985)
\newline
32. E.D'Hoker and L.Vinet, Phys. Lett. B 137, 72 (1984).
\newline
33. S.Bellucci, S.Krivonos and V.Ohanyan, Phys.Rev. D76: 105023, (2007). 
\newline
34. P.Ranjan Giri,Mod.Phys.Lett.A23:895-904, (2008). 
\newline
35. M.Salazar-Ramirez, D.Martinez, V.D.Granados and R.D.Mota, Int.J.Theor.Phys. 49 967 (2010).

\end{document}